\documentclass[preprintnumbers, floatfix,letterpaper,aps,prd,epsfig,nofootinbib,twocolumn]{revtex4-1}
\usepackage{bm,graphicx,dcolumn,epstopdf,epsf, latexsym,mathbbol, amssymb,amsmath,color,slashed, mathrsfs,mathcomp,simplewick}
\usepackage[center]{subfigure}
\usepackage{multirow}
\usepackage{makecell}
\usepackage{longtable}
\usepackage[colorlinks,linkcolor=blue,citecolor=blue,urlcolor=blue]{hyperref}

\begin{document}
\allowdisplaybreaks
 \newcommand{\bq}{\begin{equation}}
 \newcommand{\eq}{\end{equation}}
 \newcommand{\bqn}{\begin{eqnarray}}
 \newcommand{\eqn}{\end{eqnarray}}
 \newcommand{\nb}{\nonumber}
 \newcommand{\lb}{\label}
 \newcommand{\f}{\frac}
 \newcommand{\p}{\partial}
\newcommand{\PRL}{Phys. Rev. Lett.}
\newcommand{\PLB}{Phys. Lett. B}
\newcommand{\PRD}{Phys. Rev. D}
\newcommand{\CQG}{Class. Quantum Grav.}
\newcommand{\JCAP}{J. Cosmol. Astropart. Phys.}
\newcommand{\JHEP}{J. High. Energy. Phys.}
\newcommand{\red}{\textcolor{red}}
%

\title{Are the black hole remnants produced from binary black hole mergers in GWTC-3 thermodynamically stable? }

\author{Qiang Wu${}^{a, b}$}
\author{Shao-Wen Wei${}^{c,d,e}$}
\author{Tao Zhu${}^{a, b}$}
\email{corresponding author: zhut05@zjut.edu.cn}

\affiliation{
${}^{a}$ Institute for Theoretical Physics and Cosmology, Zhejiang University of Technology, Hangzhou, 310032, China\\
${}^{b}$ United Center for Gravitational Wave Physics (UCGWP), Zhejiang University of Technology, Hangzhou, 310032, China\\
${}^{c}$ Lanzhou Center for Theoretical Physics, Key Laboratory of Theoretical Physics of Gansu Province, Lanzhou 730000, China\\
${}^{d}$ School of Physical Science and Technology, Lanzhou University, Lanzhou 730000, China\\
${}^{e}$ Institute of Theoretical Physics \& Research Center of Gravitation, Lanzhou University, Lanzhou 730000, China}

\date{\today}

\begin{abstract}

Black hole thermodynamics has brought strong hints of a profound and fundamental connection between gravity, thermodynamics, and quantum theory. If the black hole does behave like a natural thermodynamic system, it should be thermodynamically stable in a clean environment. In this paper, using the observational data of binary black hole (BBH) mergers observed by LIGO, Virgo, and KAGRA detectors, we check whether the black hole remnants produced from BBH mergers in the LIGO-Virgo-KAGRA catalog GWTC-3 are thermodynamically stable. The criterion for the thermodynamic stability is quite simple and is directly related to the black hole's spin, which states that a thermodynamically stable black hole remnant requires its dimensionless spin $a>a_* \simeq 0.68$. We check the posterior distributions of final spin $a_f$ for 83 black hole remnants in GWTC-3 and find the whole remnant population is consistent with the thermodynamically stable black hole with $99.98\%$ probability. This is the first verification of the thermodynamic stability of black hole remnants produced from BBH mergers.

\end{abstract}


\maketitle

\section{Introduction}
\renewcommand{\theequation}{1.\arabic{equation}} \setcounter{equation}{0}

Since the establishment of four laws, thermodynamics has been one of the remarkable features of black holes \cite{Bekenstein:1972tm, Carlip:2014pma, Bekenstein:1973ur, Hawking:1974rv, Hawking:1975vcx}. As expected, the four laws of black hole thermodynamics involving many fundamental constants are essential keys to understanding quantum gravity. In addition, the Hawking-Page phase transition of black holes in asymptotically anti-de Sitter (AdS) space \cite{Hawking:1982dh} leads to new insights into the thermodynamics and microstructure of black holes. It has been revealed that this phase transition has an essential role in the AdS/CFT correspondence of string theory \cite{Witten:1998qj,Carlip:2014pma} and is related to the phenomena of confinement/deconfinement transitions at finite temperature in various gauge theories \cite{Mateos:2006nu, Mateos:2007vn}. Recent studies on the black hole phase transition and statistical origins of the entropy \cite{Kubiznak2012,Ruppeiner} also help us peek into the underlying microstructure of black holes \cite{ Wei:2015iwa,Wei:2019uqg}. Thus, the black hole thermodynamics is an essential constituent for exploring many open problems in modern theoretical physics.

Although black hole thermodynamics is extensively accepted, the observational evidence is quite dim. The main reason is that the Hawking temperature of a black hole is extremely weak. For example, for a Schwarzschild black hole of the solar mass, the temperature of the Hawking radiation is of the order \cite{Hawking:1974rv}
\begin{eqnarray}
T \sim 10^{-8} K,
\end{eqnarray}
which seems impossible to be directly detected by modern techniques. Recently, the advent of gravitational wave (GW) astronomy provided a new tool to explore the properties of black holes in strong-field regimes, including the black hole thermodynamics. After a hundred years of being predicted, GW was first directly detected in 2015 by LIGO \cite{gw150914}. Following this observation, about 90 GW events have been identified so far by the LIGO/Virgo/KAGRA scientific collaboration \cite{gwtc1, gwtc2,gwtc3}. With these GW events, Hawking's black hole area law \cite{Isi:2020tac} and Bekenstein-Hod universal bound \cite{Carullo:2021yxh} have been tested. While Hawking's area law is directly related to the second law of black hole thermodynamics \cite{Wald:1999vt}, the Bekenstein-Hod universal bound connects the information theory and black hole thermodynamics \cite{Hod:2006jw}. These tests thus greatly favor black hole thermodynamics. Recently, it has been proposed that the thermodynamic properties of black holes can also be used to infer the astrophysical information of BBH mergers \cite{Hu:2021lbt}.

On the other hand, it is well known that it is thermodynamically unstable for an isolated Schwarzschild black hole since its heat capacity is always negative \cite{Carlip:2014pma}. If we believe that thermodynamics plays a role in producing a black hole under a very clean environment, then it is reasonable that a Schwarzschild black hole is impossible to be the settled down state of the event while a spinning Kerr black hole may be possible. This perspective gives us another direct clue to test the black hole thermodynamics by checking its thermodynamic stability.

For a Kerr black hole, the thermodynamic stability is sensitively related to its spin. If the remnant black holes of BBH mergers are thermodynamically stable, then their dimensionless spins have to fall in the thermodynamically stable region i.e., $a_f >\sqrt{2\sqrt{3}-3}$ (c.f. Eq.~(\ref{spin})). For GW events observed by LIGO-Virgo-KAGRA, the final spin of the black hole remnants can be extracted from their GW signals. With such measurements, we present a new observational check  whether the thermodynamic stability is robust for the black hole remnants of BBH mergers in the LIGO-Virgo-KAGRA catalog GWTC-3. We find that thermodynamic stability is robust for 83 black hole remnants in GWTC-3 with 99.98\% probability. This is the first verification of the thermodynamic stability of black hole remnants produced from BBH mergers.

\section{Thermodynamics of Kerr black hole}
\renewcommand{\theequation}{2.\arabic{equation}} \setcounter{equation}{0}
\label{null}

It is expected that the spacetime of an astrophysical black hole is described by the Kerr geometry, as predicted by general relativity. For a Kerr black hole, the fundamental thermodynamic relation, which contains all the information about the thermodynamic state of a black hole, is given by the Smarr formula \cite{Smarr:1972kt},
\begin{eqnarray}
M^2 = \frac{S}{4 \pi} + \frac{\pi J^2}{S},
\end{eqnarray}
where $M$ is the mass, $S$ the entropy, and $J$ the angular momentum of the Kerr black hole. Then the first law of thermodynamics, which states the conservation of the total energy of a black hole in any thermodynamic change, takes the form
\begin{eqnarray}
dM = T dS + \Omega dJ,
\end{eqnarray}
where the temperature of the Hawking radiation $T$ and the angular velocity $\Omega$ of the event horizon are defined by
\begin{eqnarray}
T &=& \frac{\partial M}{\partial S} = \frac{1}{M} \left(1- \frac{J^2}{16 S^2}\right),\\
\Omega &=& \frac{\partial M}{\partial J} = \frac{J}{8 M S}.
\end{eqnarray}
The outer horizon radii of the Kerr black hole reads
\begin{eqnarray}
 r_{+}=(1+\sqrt{1-a^{2}})M,
\end{eqnarray}
where $a=J/M^2$ is the dimensionless black hole spin. When $a$=0, it reduces to the Schwarzschild black hole case. Obviously, the black hole horizon exists for $a\leq$1, while absent for $a>1$ corresponding to a naked singularity. In this paper we only concentrate on the black hole case corresponding to $0<a<1$.

As a thermodynamic system, it is of great importance to consider the properties of its heat capacity. For Kerr black hole, its heat capacity at constant $J$ is calculated as
\begin{eqnarray}
 C_{J}&=&T\left(\frac{\partial S}{\partial T}\right)_{J}=\frac{32\pi^4J^4S-2S^5}{-48\pi^4J^4-24\pi^2J^2S^2+S^4}.\nonumber \\
 \lb{cjv}
\end{eqnarray}
For a Schwarzschild black hole, $J=0$, and the heat capacity $C_J$ simply reduces to $-8\pi M^2$, which is negative definite. This is the well-known result that the Schwarzschild black hole is thermodynamically unstable. It could radiate energy through Hawking radiation and gets hotter \cite{thermodynamics}.

\begin{figure}
\center{
\includegraphics[width=8.1cm]{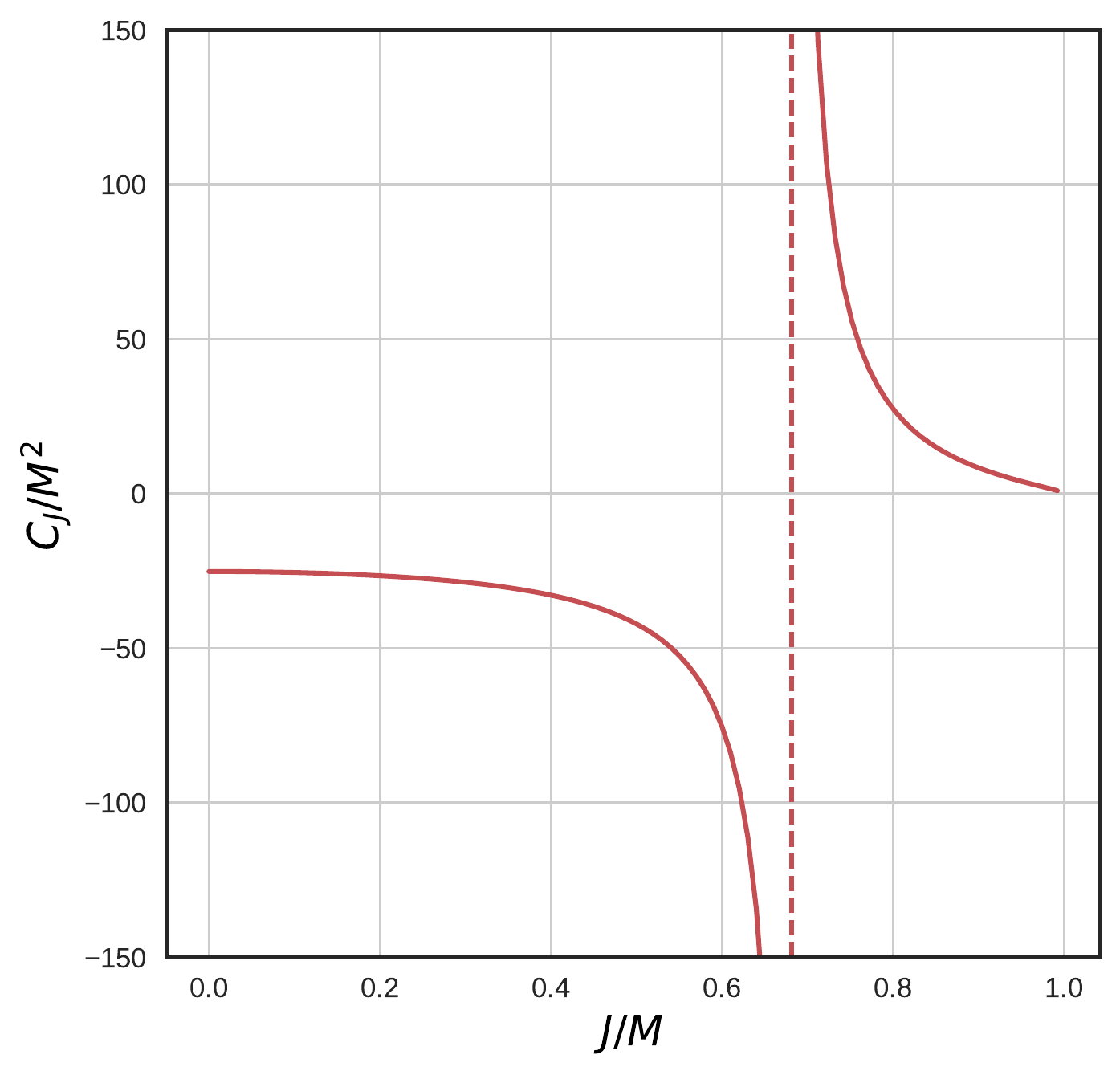}}
\caption{The behaviour of heat capacity at constant $J$. For low values of $J/M$, $C_J$ is negative, while for high values it is positive. The dashed line denotes the position of critical values $a_*=\sqrt{2 \sqrt{3}-3}$, at which the heat capacity suffers infinite discontinuity. Such discontinuity also indicates a thermodynamic phase transition at $a_*$. Note that we set $M=1$ in the plot.}\label{CJ}
\end{figure}

The behavior of the heat capacity $C_J$ for a Kerr black hole is illustrated in Fig.~\ref{CJ}. It is obvious that for low values of $a$, the heat capacity $C_{J}$ is negative, while for high values, it is positive. This indicates that the Kerr black hole is thermodynamically unstable for low values of $a$, and is stable for high values. One conspicuous feature in this figure is the infinite discontinuity at a specific value of $a$ at which the heat capacity $C_J$ changes from negative to positive. Such a transition also indicates a crucial thermodynamic phase transition of the Kerr black hole \cite{thermodynamics}. One can estimate the critical value of $a$ at which the discontinuity occurs by directly requiring the denominator of Eq.~(\ref{cjv}) vanishes, which yields
\begin{eqnarray}
a_* = \sqrt{2 \sqrt{3}-3}\simeq 0.68. \lb{spin}
\end{eqnarray}
{\em This critical value thus provides an absolute criterion beyond which the Kerr black hole is thermodynamically stable.}

\section {Analysis of the posterior samples of final spin}
 \renewcommand{\theequation}{3.\arabic{equation}} \setcounter{equation}{0}

It is reasonable to believe that the neighborhood of these merged black holes is quite clean, so it is an excellent opportunity to measure the final spin of the remnants of BBH merges. This provides a direct approach to check if the black hole remnants produced by the BBH mergers are thermodynamically stable. If the spin falls in the region of thermodynamic stability, we then can conclude that black hole thermodynamics is favored.

 For our purpose, we consider the posterior samples of BBH mergers in the catalogs released recently by LIGO-Virgo-KAGRA Collaboration \cite{gwtc1, gwtc2, gwtc3}. Not all the GW events in these catalogs are of our interest. First, the compact binary coalescence events involving neutron stars would be discarded in such an analysis since they might be related to unknown matter effects rather than the pure-gravity effect. We also discard those events of the binary system with one component's mass $\lesssim 3 M_{\odot}$ because of the uncertainty of this component's property. Here $M_{\odot}$ denote the solar mass. Therefore, we consider 83 events of BBH mergers in total in the LIGO-Virgo-KAGRA catalog as described by GWTC-3. The data of the posterior samples of the final spin for these 83 GW events are downloaded from Gravitational Wave Open Science Center \cite{GWTC}.

 \begin{figure}
\includegraphics[height=20.5cm]{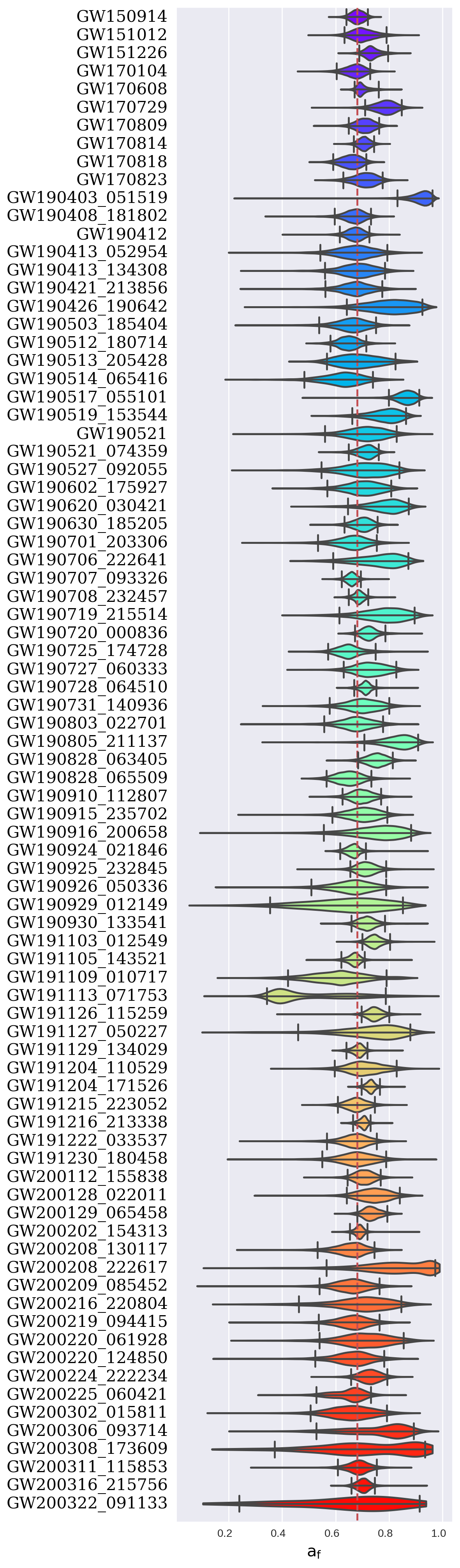}
\caption{Marginal posterior distributions on the final spin of the remnants for  83 BBH GW events in  GWTC-3. The results are directly obtained from the posterior samples released by the LIGO-Virgo-KAGRA collaboration. The left and right bar regions denote the 90\% credible interval, and the red dashed line represents the critical value of the final spin $a_{*} = \sqrt{2 \sqrt{3} -3}$.
} \label{violin}
\end{figure}

\begin{figure}
{
\includegraphics[width=8.1cm]{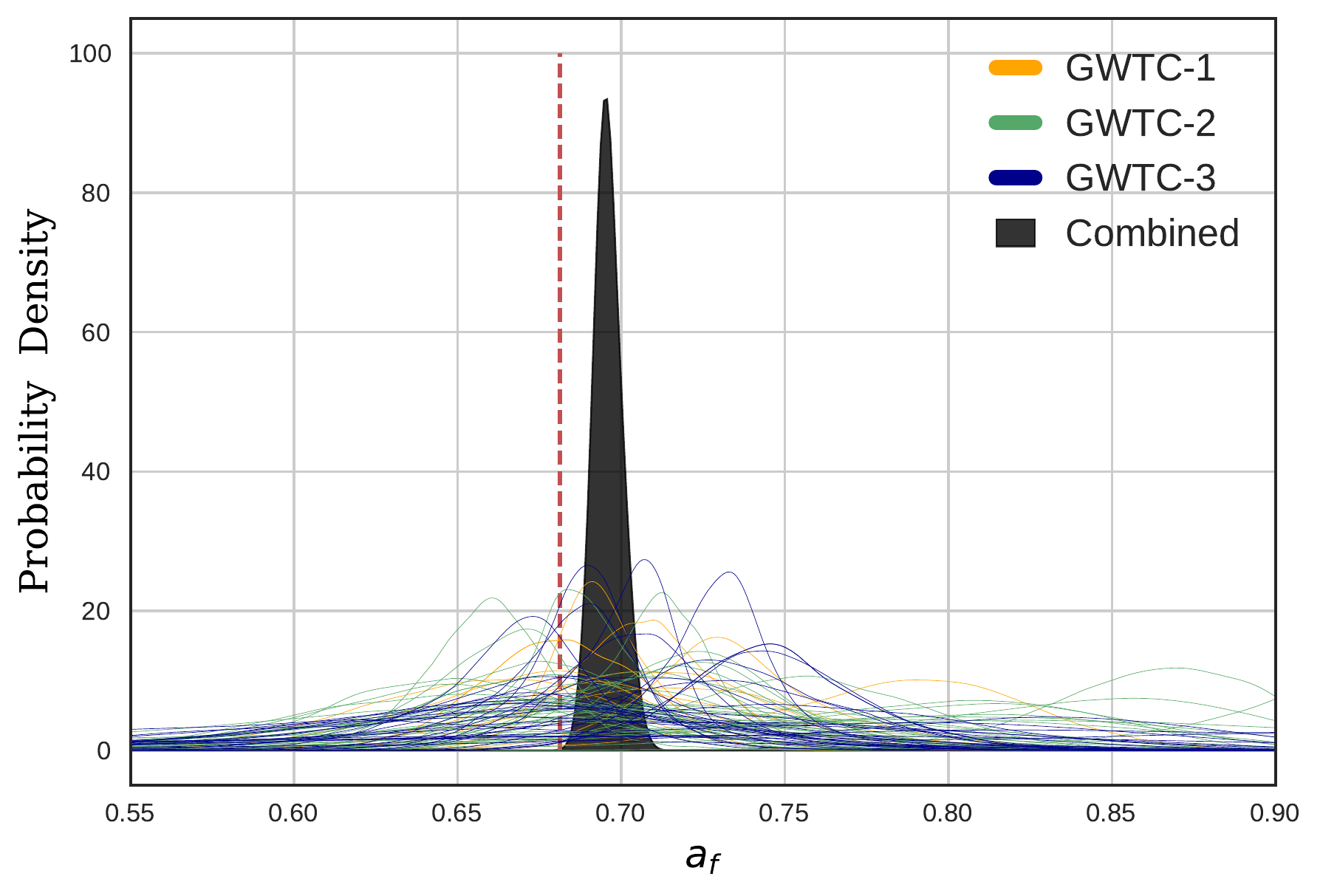}}\\
\caption{Posterior probability distributions of remnant spin $a_f$. The results are obtained by analyzing the selected 83 GW events. The combination (black area) of all the events is displayed, and the red vertical dash line indicates the spin $a_{\rm f}=a_{*} = \sqrt{2 \sqrt{3} -3}$.} \label{density}
\end{figure}

In order to verify whether the black hole remnants are thermodynamically stable, we first check the posterior distribution of the final spin $a$ for each GW event we analyzed in this paper. In Fig.~\ref{violin} we present briefly the posterior distributions of the final spin $a_f$ for 83 GW events of BBH mergers. In this figure, the left and right bar regions denote the 90\% credible interval for all these events. The red dashed line indicates the position of the critical value $a_*$. From Fig.~\ref{violin}, except for a few events, one observes that most of the GW events we analyzed are consistent with $a>a_*$ at certain probabilities. We also plot the posterior probability distributions of remnant spin $a_f$ in Fig.~\ref{density}. From this figure, we find that the median values with 90\% confidence intervals for most of GW events analyzed are consistent with the criterion of thermodynamic stability, i.e., $a_f>a_* \simeq 0.68$, with high probabilities. A few events, although their probabilities are a little bit lower, are still consistent with the criterion of thermodynamic stability. These results show that most of the events we considered in our analysis are highly in agreement with the thermodynamic stability of the black hole remnants. No significant disagreement with the thermodynamic stability of the remnant is found.

The final spin for each black hole remnant depends on the source parameters of the GW signal. Thus, different remnants have different posterior distributions, which are denoted by ${P(a_f|d_i)}$ where $d_i$ denotes data of the $i$th GW event. Considering the thermodynamic properties of black holes are universal, it is expected that the posterior distributions $P(a_f|d_i)$ for all the black hole remnants will show significant supports of $a_f >a_*$ if these remnants are thermodynamically stable. For this reason, we can consider the stability criterion as a universal property for all the black hole remnants. Then we can extract the combined probability distribution density of the final spin parameter by multiplying the individual posterior distributions through
\begin{eqnarray}
P(a_f) \propto \prod_{i=1}^{N} P(a_f| d_i),
\end{eqnarray}
where $N$ is the total number of GW events we analyzed. To achieve this, we applied the Gaussian kernel density estimation to smooth the histogram using the method proposed in \cite{Yunes}. The results are presented in Fig.~\ref{density} from which we find that an overall probability $P(a_f>a_*)$ of the whole population of BBH remnants falls in the region of thermodynamically stable to be significantly high with a value of
\begin{eqnarray}
P(a_f>a_*) = 99.98\%.
\end{eqnarray}
We then conclude that the population of BH remnants produced from the 83 BBH mergers in the LIGO-Virgo-KAGRA catalog GWTC-3 are thermodynamically stable with a high probability of 99.98\%.

{\em Conclusions}--- The black hole thermodynamics is essential for exploring many open problems in modern theoretical physics. However, the direct observational evidence of the black hole thermodynamics is still quite lacking. Recent detections of GWs from BBH mergers provide novel tests of Hawking's area law \cite{Isi:2020tac} and Bekenstin-Hod universal bound \cite{Carullo:2021yxh}, which strongly suggest that black hole does follow the thermodynamic laws. In this work, we consider another natural feature of black hole thermodynamics: whether the black hole remnants produced from BBH mergers detected by LIGO-Virgo-KAGRA are thermodynamically stable. By performing an analysis on the posterior distribution of the final spin for 83 remnants in GWTC-3, we provide the first observational verification of the thermodynamic stability of black hole remnants with a significantly high probability of $99.98\%$. Our analysis offers another strong hint of the existence of the black hole thermodynamics.


{\em Acknowledgements}--- We thank Yi-Fan Wang and Rui Niu for helpful discussions. T.Z. and Q.W. are supported by the National Key Research and Development Program of China Grant No. 2020YFC2201503, and the Zhejiang Provincial Natural Science Foundation of China under Grants No. LR21A050001 and No. LY20A050002, the National Natural Science Foundation of China under Grant No. 11675143, and the Fundamental Research Funds for the Provincial Universities of Zhejiang in China under Grant No. RF-A2019015. S.W.W. is supported by the National Natural Science Foundation of China under Grants No. 12075103 and No. 11675064, and No. 12047501.


\begin{thebibliography}{399}

\bibitem{Bekenstein:1972tm}
J.~D.~Bekenstein,
Black holes and the second law,
Lett. Nuovo Cim. \textbf{4}, 737 (1972).

\bibitem{Bekenstein:1973ur}
J.~D.~Bekenstein,
Black holes and entropy,
Phys. Rev. D \textbf{7}, 2333 (1973).

\bibitem{Hawking:1974rv}
S.~W.~Hawking,
Black hole explosions,
Nature \textbf{248}, 30 (1974).

\bibitem{Hawking:1975vcx}
S.~W.~Hawking,
Particle Creation by Black Holes,
Commun. Math. Phys. \textbf{43}, 199 (1975)
[erratum: Commun. Math. Phys. \textbf{46}, 206 (1976)].

\bibitem{Carlip:2014pma}
S.~Carlip,
Black Hole Thermodynamics,
Int. J. Mod. Phys. D \textbf{23}, 1430023 (2014).

\bibitem{Hawking:1982dh}
S.~W.~Hawking and D.~N.~Page,
Thermodynamics of Black Holes in anti-De Sitter Space,
Commun. Math. Phys. \textbf{87}, 577 (1983).

\bibitem{Witten:1998qj}
E.~Witten,
Anti-de Sitter space and holography,
Adv. Theor. Math. Phys. \textbf{2}, 253 (1998)
[arXiv:hep-th/9802150 [hep-th]].

\bibitem{Mateos:2006nu}
D.~Mateos, R.~C.~Myers and R.~M.~Thomson,
Holographic phase transitions with fundamental matter,
Phys. Rev. Lett. \textbf{97}, 091601 (2006)
[arXiv:hep-th/0605046 [hep-th]].

\bibitem{Mateos:2007vn}
D.~Mateos, R.~C.~Myers and R.~M.~Thomson,
Thermodynamics of the brane,
JHEP \textbf{05}, 067 (2007)
[arXiv:hep-th/0701132 [hep-th]].

\bibitem{Kubiznak2012}
D.~Kubiznak and R.~B.~Mann,
$P$-$V$ criticality of charged AdS black holes,
JHEP \textbf{07}, 033 (2012)
[arXiv:1205.0559 [hep-th]].

\bibitem{Ruppeiner}
G.~Ruppeiner ,
Thermodynamic curvature and black holes,
Springer Proc.Phys. \textbf{153}, 179 (2014)
[arXiv:1309.0901 [gr-qc]].

\bibitem{Wei:2015iwa}
S.~W.~Wei and Y.~X.~Liu,
Insight into the Microscopic Structure of an AdS Black Hole from a Thermodynamical Phase Transition,
Phys. Rev. Lett. \textbf{115}, 111302 (2015)
[erratum: Phys. Rev. Lett. \textbf{116}, 169903 (2016)].

\bibitem{Wei:2019uqg}
S.~W.~Wei, Y.~X.~Liu and R.~B.~Mann,
Repulsive Interactions and Universal Properties of Charged Anti\textendash{}de Sitter Black Hole Microstructures,
Phys. Rev. Lett. \textbf{123}, 071103 (2019).

\bibitem{gw150914}
B. P. Abbott {\em et al.} (LIGO Scientific and Virgo Collaborations),
Observation of Gravitational Waves from a Binary Black Hole Merger,
Phys. Rev. Lett. {\bf 116}, 061102 (2016)
GW150914: First results from the search for binary black hole coalescence with Advanced LIGO,
Phys. Rev. D {\bf 93}, 122003 (2016)
Properties of the Binary Black Hole Merger GW150914,
Phys. Rev. Lett. {\bf 116}, 241102 (2016);
GW150914: The Advanced LIGO Detectors in the Era of First Discoveries,
Phys. Rev. Lett. {\bf 116}, 131103 (2016).


\bibitem{gwtc1}
B.~P.~Abbott \textit{et al.} (LIGO Scientific and Virgo Collaborations),
``GWTC-1: A Gravitational-Wave Transient Catalog of Compact Binary Mergers Observed by LIGO and Virgo during the First and Second Observing Runs,''
Phys. Rev. X \textbf{9}, 031040 (2019)
[arXiv:1811.12907 [astro-ph.HE]].

\bibitem{gwtc2}
R.~Abbott \textit{et al.} (LIGO Scientific and Virgo Collaborations),
``GWTC-2: Compact Binary Coalescences Observed by LIGO and Virgo During the First Half of the Third Observing Run,''
Phys. Rev. X \textbf{11}, 021053 (2021)
[arXiv:2010.14527 [gr-qc]].

\bibitem{gwtc3}
R.~Abbott \textit{et al.} (LIGO Scientific, VIRGO, and KAGRA Collaborations),
GWTC-3: Compact Binary Coalescences Observed by LIGO and Virgo During the Second Part of the Third Observing Run,
arXiv:2111.03606 [gr-qc].


\bibitem{Isi:2020tac}
M.~Isi, W.~M.~Farr, M.~Giesler, M.~A.~Scheel, and S.~A.~Teukolsky,
Testing the Black-Hole Area Law with GW150914,
Phys. Rev. Lett. \textbf{127}, 011103 (2021)
[arXiv:2012.04486 [gr-qc]].

\bibitem{Carullo:2021yxh}
G.~Carullo, D.~Laghi, J.~Veitch, and W.~Del Pozzo,
Bekenstein-Hod Universal Bound on Information Emission Rate Is Obeyed by LIGO-Virgo Binary Black Hole Remnants,
Phys. Rev. Lett. \textbf{126}, 161102 (2021)
[arXiv:2103.06167 [gr-qc]].

\bibitem{Wald:1999vt}
R.~M.~Wald,
The thermodynamics of black holes,
Living Rev. Rel. \textbf{4}, 6 (2001)
[arXiv:gr-qc/9912119 [gr-qc]].

\bibitem{Hod:2006jw}
S.~Hod,
Universal Bound on Dynamical Relaxation Times and Black-Hole Quasinormal Ringing,
Phys. Rev. D \textbf{75}, 064013 (2007)
[arXiv:gr-qc/0611004 [gr-qc]].

\bibitem{Hu:2021lbt}
P.~Hu, K.~Jani, K.~Holley-Bockelmann and G.~Carullo,
Thermodynamics to infer the astrophysics of binary black hole mergers,
arXiv:2112.06856 [gr-qc].

\bibitem{Smarr:1972kt}
L.~Smarr,
Mass formula for Kerr black holes,
Phys. Rev. Lett. \textbf{30}, 71 (1973)
[erratum: Phys. Rev. Lett. \textbf{30}, 521 (1973)].

\bibitem{thermodynamics}
P.~C.~W. Davies,
The thermodynamic theory of black holes,
Proc. R. Soc. Lond. A. \textbf{353}, 499 (1977).

\bibitem{GWTC}
The Gravitational Wave Open Science Center: \href{https://www.gw-openscience.org/}{https://www.gw-openscience.org/}

\bibitem{Yunes}
S. E. Perkins, R. Nair, H.O. Silva, and N. Yunes,
Improved gravitational-wave constraints on higher-order curvature theories of gravity,
Phys. Rev. D \textbf{104}, 024060 (2021).

\end{thebibliography}
\end{document}